\documentclass[twocolumn]{article}
\usepackage[margin=0.6in,columnsep=0.15in]{geometry}
\usepackage[utf8]{inputenc}

\usepackage[usenames,dvipsnames]{xcolor}
\usepackage[colorlinks,citecolor=Gray,urlcolor=Gray,linkcolor=Blue]{hyperref}

\usepackage{graphicx}
\usepackage{mathtools}
\usepackage{booktabs}
\usepackage{tabularx}
\usepackage{tikz}
\usepackage{siunitx}
\usepackage[normalem]{ulem}
\usepackage[lcgreekalpha,upint]{stix}

\usepackage[tiny]{titlesec}
\usepackage[inline]{enumitem}
\usepackage[normal]{caption}
\DeclareCaptionLabelSeparator{pipe}{ $|$ }
\usepackage{float}
\usepackage{authblk}
\usepackage[square,sort&compress,comma,numbers]{natbib}

\usepackage{subfigure}

\usepackage{hyperref}

\usepackage[switch]{lineno}
\title{\Large Emergent Wigner phases in moir\'e superlattice from deep learning}

\author[1*]{Xiang Li}
\author[1,4]{Yubing Qian}
\author[1]{Weiluo Ren}
\author[2,3$\dagger$]{Yang Xu}
\author[4,5$\ddagger$]{Ji Chen}
\affil[1]{ByteDance Research, Fangheng Fashion Center, No. 27, North 3rd Ring West Road, Haidian District, Beijing 100098, People’s Republic of China}
\affil[2]{Beijing National Laboratory for Condensed Matter Physics, Institute of Physics, Chinese Academy of Sciences, Beijing 100190, China}
\affil[3]{School of Physical Sciences, University of Chinese Academy of Sciences, Beijing 10049, China}
\affil[4]{School of Physics, Peking University, Beijing 100871, People’s Republic of China.}
\affil[5]{Interdisciplinary Institute of Light-Element Quantum Materials, Frontiers
Science Center for Nano-Optoelectronics, Peking University, Beijing 100871, People’s Republic of China}

\date{}

\makeatletter
\def\blfootnote{\xdef\@thefnmark{}\@footnotetext}
\makeatother
\date{\today}% It is always \today, today,
\begin{document}

\twocolumn[{%
  \maketitle
  \vspace{-3em}
  \begin{center}
  \begin{minipage}{0.85\linewidth}
    \small
    \paragraph{Abstract}
    Moir\'e superlattice designed in stacked van der Waals material provides a dynamic platform for hosting exotic and emergent condensed matter phenomena. 
    %One particularly attractive aspect of moir\'e system is the strong electron correlation physics. 
    %(Due to the large size of moire unit cell and computation limit, one particular obstacle lies in predicting new electronic phases.) 
    %
    However, the relevance of strong correlation effects and the large size of moir\'e unit cells pose significant challenges for traditional computational techniques.
    %in predicting new electronic phases.
    %
    To overcome these challenges, we develop an unsupervised deep learning approach to uncover electronic phases emerging from moir\'e systems based on variational optimization of neural network many-body wavefunction.
    %Here we develop a deep learning approach to discover electronic phases emerging from moir\'e systems. Based on the variational principle, neural network many-body wavefunction is trained in an unsupervised manner. 
    %
    % Using our approach, various states with different fillings and spins are investigated, and an abundance of quantum states emerge from our network, including exotic phases of generalized Wigner crystals, Wigner molecular crystals, and unreported Wigner covalent crystals. 
    Our approach has identified diverse quantum states, including novel phases such as generalized Wigner crystals, Wigner molecular crystals, and previously unreported Wigner covalent crystals. These discoveries provide insights into recent experimental studies and suggest new phases for future exploration. 
    %
    %
    %(spin orientation plays crucial rule for determining the ground-state charge distribution)
    %The electronic behavior of the system also varies due to different spin polarizations, and can be manipulated using external magnetic fields.
    %
    %Our results not only shed light on the recent experimental reports and discover unexplored phases for further investigation, but also highlight the crucial role of spin polarization in determining Wigner phases. 
    % Our results shed light on the recent experimental reports and discover unexplored phases for further investigation. 
    %
    They also highlight the crucial role of spin polarization in determining Wigner phases. 
    %
    %More importantly, we propose a general approach to study moir\'e physics with deep learning technique.
    More importantly, our proposed deep learning approach is proven general and efficient, offering a powerful framework for studying moir\'e physics.
    
  \end{minipage}
  \end{center}
  \vspace{1em}
}]

\blfootnote{$^*$ lixiang.62770689@bytedance.com}%
\blfootnote{$^\dagger$ yang.xu@iphy.ac.cn}%
\blfootnote{$^\ddagger$ ji.chen@pku.edu.cn}%

\maketitle
% \twocolumngrid
\paragraph{Main}
%Moir\'e superlattice has received extensive investigation in the past few years, starting with twisted graphene and quickly generalized to transition metal dichalcogenide (TMD) and other systems. Through experimental manipulation, their electronic band structure can be artificially controlled and driven to strong correlation region, which has remained a mystery in condensed matter physics for decades. Novel quantum phases of distinct symmetry and particle occupancy, such as the generalized Wigner crystal, have been detected in laboratories recently, providing significant understanding into the strong correlation phenomenon.
\begin{figure*}[t!]
\centering
\includegraphics[width=2\columnwidth]{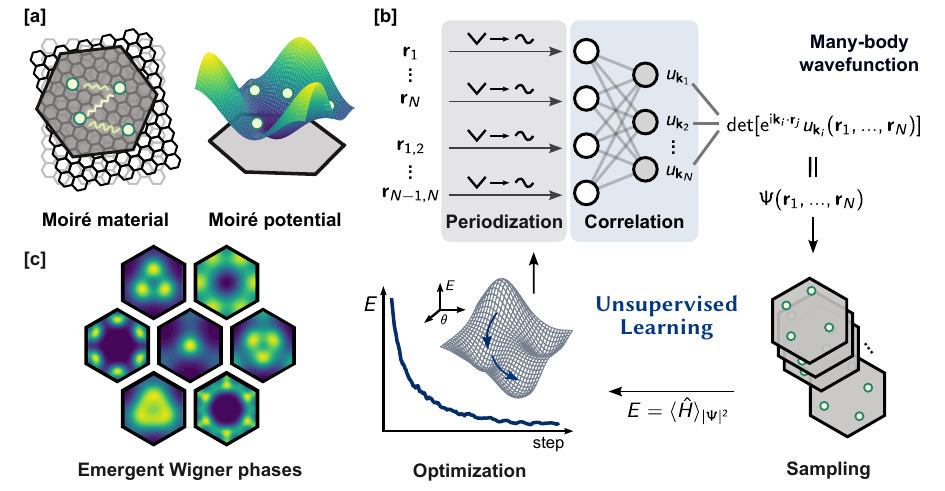}
\caption{\label{fig:workflow} \textbf{Workflow of deep learning approach}. \textbf{a}, plotted structure of moir\'e materials, particles are constrained by moir\'e potential and interact with each other via Coulomb interaction. \textbf{b}, a sketch of deep learning approach. 
%A neural network is built for moir\'e superlattice.
The upper part illustrates a neural network wavefunction for a moir\'e superlattice.
Particle features $\mathbf{r}$ are converted to be periodic and then fed into neural networks, forming the many-body wavefunction \textit{ansatz} $\Psi$. 
%
%Variational Monte Carlo is then applied to sample patterns and optimize the energy expectations $\langle\hat{H}\rangle_{|\Psi|^2}$, reaching the ground state. 
The neural network is then trained using unsupervised learning.
Specifically, variational Monte Carlo is applied to sample patterns and minimize the energy expectation $\langle\hat{H}\rangle_{|\Psi|^2}$.
During the training iterations, the neural network wavefunction gradually approaches the ground state.
\textbf{c}, emergent Wigner crystal phases in our simulation, including Mott insulator, Wigner molecular crystals, and other exotic phases. }
\end{figure*}

Moir\'e superlattices, spanning from twisted graphene \cite{bilayer_graphene_theory,bilayer_graphene_exp, bilayer_graphene_exp2}, transition metal dichalcogenides (TMDs) \cite{TMD_model, wse2_hubbard, fractional_exp1, fractional_exp2, fractional_exp3, fractional_exp4}, and other systems, have attracted significant research interest in condensed matter physics over the last few years \cite{andrei_rev}.
They provide a flexible platform to tune the electronic, magnetic, and optical properties of materials, and explore strongly correlated and topological phenomena such as unconventional superconductivity \cite{bilayer_graphene_exp, TBG_sc} and correlated insulating behaviors \cite{bilayer_graphene_exp,bilayer_graphene_exp2, TMD_model,fractional_exp1, fractional_exp2, fractional_exp3, fractional_exp4, wse2_hubbard}. 
For example, recent discoveries of new quantum phases, including the generalized Wigner crystal \cite{fractional_exp1, fractional_exp2, fractional_exp3, fractional_exp4}, known for their unique symmetry and particle arrangements, have provided deep insights into the behaviors of strongly correlated systems. 
These insights have been further enriched recently by Wigner molecular crystals emerging from multi-hole artificial atoms in twisted ${\rm WS_2}$ homobilayer \cite{wigner_molcule_theory, wigner_molecule_exp}, underscoring the tunable feature of moir\'e superlattices.
%and offering novel platforms for investigating quantum phenomena.

%In order to comprehend these emerging quantum phases, different theoretical methods have been proposed, including classical Monte Carlo simulation, effective Hubbard model, and Hartree–Fock calculation. These techniques offer a compelling explanation for observed quantum phases, and successfully predict additional unidentified phases in experiments. 

% Computational methods like classical Monte Carlo simulation and self-consistent Hartree–Fock method have been utilized to understand these systems. 
% %
% However, accurate calculations for moiré superlattice face significant challenges due to the strong correlation effects and the large system sizes required for reliable simulations.
% %
% Yet, high-accuracy methods are still in need to incorporate more quantum correlations.
% %
%%
Theoretically, a grand challenge remains that a general theoretical approach to handling strong electron correlation effects of moir\'e system is lacking,
and the current investigations mostly rely on traditional computational methods, such as classic Monte Carlo \cite{fractional_exp1} and Hartree-Fock approximation \cite{HF, wigner_molcule_theory,  wigner_molecule_exp}. 
While these calculations may provide qualitative explanations for experimental results and make other predictions, it is questionable whether all essential correlated physics has been captured.
The importance of electron correlation effects on the phase transitions in a moiré system is further highlighted with a combination of auxiliary field quantum Monte Carlo and diffusion Monte Carlo \cite{vmc_moire}. 
%
%but also encounter challenges to deal with strong correlation effects, necessitating a high-accuracy method to incorporate more quantum correlations.
%
%Nonetheless, due to the moir\'e system's strong correlation character, a high-accuracy method is essential to achieve a general and thorough study of these systems. However, precision often accompanies expensive computational scaling rendering the simulation prohibitively costly. This dilemma has somehow changed in these years, after neural network-based method is proposed in electronic structure community. Specifically speaking, neural network is proposed as a wavefunction \textit{ansatz} in simulation, which combined with quantum Monte Carlo (QMC) method, achieving competitive accuracy compared to traditional method with an economical computation scaling.
In recent years, powerful approaches based on deep learning architecture have been developed to treat quantum many-body problems.
Based on the universality and expressiveness of neural networks, quantum many-body states can be well-represented without assuming a limited traditional \textit{ansatz}, and new insights into the correlation effects are revealed.
However, so far these deep learning approaches can only be applied to simple lattice models \cite{rbm}, chemical molecules \cite{deephf,ferminet, paulinet},  uniform electron gases \cite{deepsolid, wapnet, ferminet_heg}, small solids \cite{deepsolid}, and isolated Wigner molecules \cite{wigner_molecule_net}.
\begin{figure*}[t]
\centering
   \includegraphics[width=2\columnwidth]{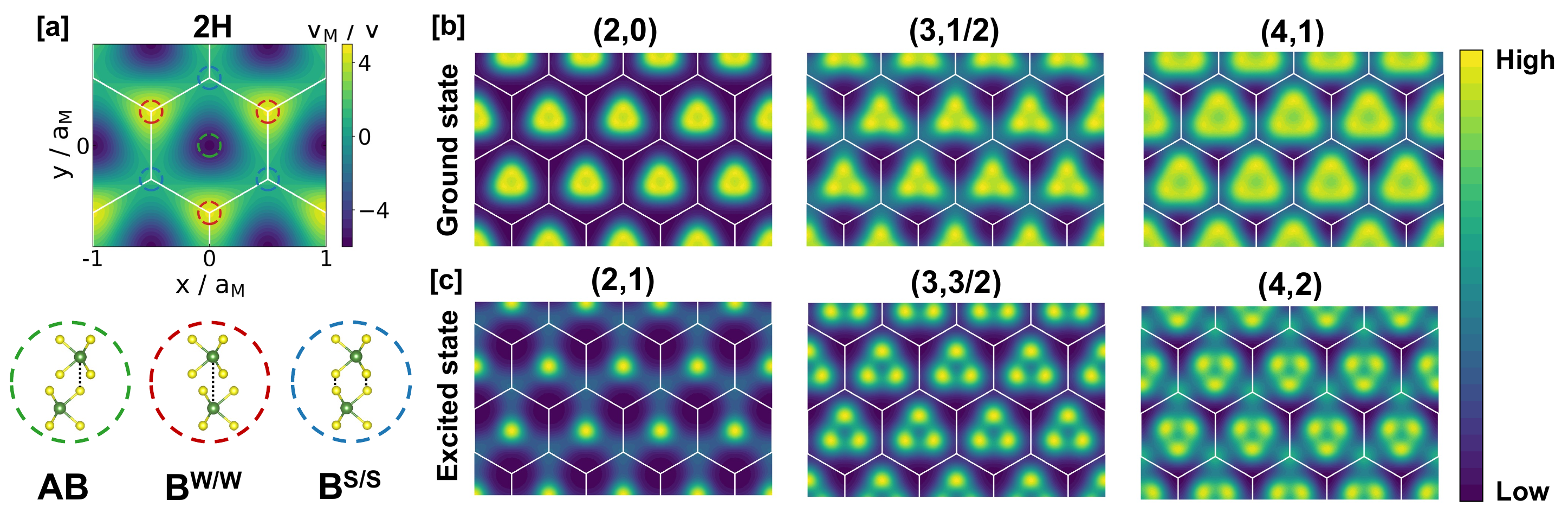}
    \caption{\textbf{Calculated density pattern of ${\rm 2H\ WS_2}$ homobilayer}. Tuples above the figure denote the particle fillings and the magnetic number of spin ($\nu$, $S_z$). \textbf{a}, moir\'e potential of ${\rm 2H\ WS_2}$ homobilayer. $a_M$ denotes length of moir\'e cell and is set to $9.8\ {\rm nm}$. Colored circles denote regions with different stack patterns. \textbf{b-c}, calculated ground state and excited state of ${\rm 2H\ WS_2}$ at different particle fillings $\nu$ and magnetic numbers $S_z$.}
    \label{fig:2h}
\end{figure*}
%
%
%this approach has been successfully applied periodic models, real solids and other systems \cite{deepsolid, wapnet, ferminet_heg, wigner_molecule_net}, demonstrating its ability to achieve highly-competitive accuracy.
%
%In this work, neural network-based wavefunction method is extended to moir\'e physics. TMD materials of various component are simulated at different particle fillings. Interesting phase, including generalized Wigner crystal, Mott insulator, and recent reported Wigner molecular crystal, appears sequentially as we tune the particle fillings. Candidate of undiscovered quantum phases are also given in this work for future studies. 
%

In this work, we develop a neural network based wavefunction methodology for moir\'e systems.
The neural network wavefunction is trained via the variational quantum Monte Carlo method in an unsupervised manner.
Focusing on the family of TMD materials over a wide range of particle fillings,
we unveil a sequence of intriguing phases.
%highlighting the influence of particle fillings on the system's property.
At fractional fillings, symmetry preserved and broken Wigner crystal emerges with varying fillings.
At integer fillings, electron correlations lead to a rich diagram of Wigner phases, including the recently observed Wigner molecular crystals, and Wigner covalent crystals discovered in this work.
%
%Furthermore, we predict additional novel quantum phases, pending validation through future experiments.
%
%This study not only forges a link between theoretical models and experimental realities but also 
This study underscores the pivotal role of electron correlations captured by deep learning in unlocking and leveraging the quantum phenomena inherent in moir\'e superlattices.

\paragraph{Deep learning architecture}
Fig.~\ref{fig:workflow} summarizes the overall workflow.
First of all, to explore the quantum phases of moir\'e materials, a neural network is employed to represent the many-body wavefunction. 
The neural network wavefunction $\Psi_{\rm net}$ has the following form
\begin{equation}
\begin{gathered}
\Psi_{\rm net}=\det[e^{i\mathbf{k}_i\cdot \mathbf{r}_j}u_{\mathbf{k}_i}(\mathbf{r}_1,...,\mathbf{r}_N)]\ .
\end{gathered}
\label{eq:net}
\end{equation}
Compared with traditional wavefunction, $\Psi_{\rm net}$ explicitly includes the features of all particles $(\mathbf{r}_1,...,\mathbf{r}_N)$ in the formulation, permitting it to capture all the correlations. Specifically, particle features $\mathbf{r}_i$ are converted to be periodic and fed into permutation-equivalent neural networks \cite{ferminet, deepsolid} to construct the generalized Bloch functions $u_{\mathbf{k}_i}(\mathbf{r}_1,...,\mathbf{r}_N)$ in Eq.~\eqref{eq:net}, which together with the Slater determinant form a periodic, complex-valued, and antisymmetric many-body wavefunction for superlattice. Millions of parameters are embedded in the neural network, making it a powerful and convenient tool for expressing quantum states of distinct symmetries. 

Given the neural network wavefunction and the Hamiltonian of a real material, one can employ the variational Monte Carlo (VMC) method to efficiently optimize the neural network parameters. VMC is based on the variational principle of quantum mechanics, which states the ground state has the lowest energy among all the solutions to the stationary Schr\"{o}dinger equation. Therefore, the VMC training of the neural network is an unsupervised process, which further guarantees the reliability of our deep learning approach.

\paragraph{Moir\'e superlattice} Moir\'e superlattice appears when two layers of van der Waals material overlap each other, with mismatched lattice constants and/or twisted angles. The superlattice formed features a significantly large lattice constant and contains thousands of atoms in the unit cell, which makes it unrealistic to employ the full \textit{ab initio} Hamiltonian of a moir\'e material in the current workflow. Therefore, in this work we focus on the low-energy effective Hamiltonian of TMDs \cite{TMD_model}, which reads
\begin{equation}
\begin{gathered}
\hat{H}=\sum_i\left[-\frac{\Delta_i}{2m^*}+V_M(\mathbf{r}_i)\right]+\frac{1}{2}\sum_{i\neq j}\frac{1}{\epsilon|\mathbf{r}_i-\mathbf{r}_j|}\ , \\
V_M(\mathbf{r})=-2V\sum_{i=1}^3 \cos(\mathbf{b}_i\cdot\mathbf{r}+\phi)\ . 
\end{gathered}
\label{eq:model}
\end{equation}
This model describes doping holes of TMD materials near Fermi surface and $m^*$ refers to the effective mass of valence band edge. $V_M$ is moir\'e potential experienced by holes across the whole moir\'e materials. %$\mathbf{b}_i = \frac{4\pi}{\sqrt{3}a_M}(\sin\frac{2\pi i}{3}, \cos\frac{2\pi i }{3})$  
$\mathbf{b}_i$ denotes the reciprocal vector of moir\'e cell,
and $V,\phi$ are parameters depending on the material. When moir\'e materials get twisted to critical angles, flat band appears and Coulomb interaction becomes dominant, which accounts for the last term in Eq.~\eqref{eq:model} with $\epsilon$ referring to effective dielectric constant. A uniform charge background is also necessary to make the whole system neutral and remove divergence in Coulomb interactions.

The effective Hamiltonian has included the key feature of Coulomb interactions and greatly simplifies the problem, so it has been successfully used in studies of TMD heterobilayers such as ${\rm WSe_2/WS_2}$ and $\Gamma$-valley homobilayers such as twisted ${\rm WS_2}$ \cite{TMD_model, WS2_3R}.
It is worth noting that electron correlations in such an effective Hamiltonian are as non-trivial as the \textit{ab initio} Hamiltonian, and an exact solution remains extremely difficult. The most prevalent Hartree-Fock method treats particles independent from each other, which would lose the correlation effects. More accurate methods such as the full configuration interaction theory exist \cite{fci, fci2, hci}, but they are only applicable to isolated small molecules. Our deep learning wavefunction approach achieves an optimal balance of accuracy and efficiency, making it a very promising tool for studying moir\'e systems.

\paragraph{Generalized Wigner crystal}
%In 1934, Eugene Wigner imagined a special crystal which is composed purely of electrons. As the electron gas becomes sparse, the kinetic energy of electrons is dominated by the Coulomb potential, then electrons will spontaneously organize themselves to present spatial periodicity. The seek for Wigner crystal has continued ever since, leading to many important discoveries in condensed matter physics in the past century. In recent years, the seek has been extended to generalized Wigner crystal in moire materials \cite{fractional_exp2}. 

In 1934, Eugene Wigner imagined an electron-only crystal, where electrons spontaneously organize into special patterns as their density decreases.
The pursuit of Wigner crystals has continued ever since, and driven significant condensed matter physics discoveries for decades.
Recently, the investigation is expanded to generalized Wigner crystals in moir\'e materials \cite{fractional_exp1,fractional_exp2,fractional_exp3,stripe}. 
%
%As an example, we employ our neural network to simulate a ${\rm WSe_2/WS_2}$ heterobilayer (see Fig.~\ref{fig:frac}a) at low particle fillings $\nu\leq 1$, and the results are plotted in Fig.~\ref{fig:frac}c-d.
We first employ our deep learning methods to reproduce the experimentally observed Wigner phases 
%As an example, we employ our neural network to simulate 
in a ${\rm WSe_2/WS_2}$ heterobilayer at fractional particle fillings $\nu\leq 1$, which serves as a validation of our methodology.
The results are presented in Supplementary Fig.~4.
%The first interesting observation is that even at $\nu=1$ it is found to be a Mott insulator phase instead of metal, highlighting the importance of electron correlation in this system. Going towards lower fractional fillings, we can see various phases emerging. With $\nu=1/3,2/3$, the $C_3$ symmetry of the moire lattice is preserved, leading to commensurate Wigner crystal phases.
%
In line with recent experimental results \cite{fractional_exp1,fractional_exp2,fractional_exp3, fractional_exp4}, we observe a Mott insulator phase instead of metal at $\nu=1$ manifesting the strong correlation characters of such systems.
We also observe various Wigner crystal phases at $\nu = 1/3, 2/3$ with $C_3$ symmetry, and
symmetry-breaking stripe phases at $\nu=1/2,2/5,1/6$, 
%On the other hand, $C_3$ symmetry of the materials is found to be broken where a range of stripe phases appear, 
which are also consistent with recent optical anisotropy and electronic compressibility measurements \cite{stripe}. 
%
%Among these states, the most remarkable one is $\nu=2/5$, in which holes attempt to attract each other despite Coulomb repulsion, forming a Cooper-pair-like object and leading to possible superconductivity in Mott insulators \cite{andrei_rev}.
%
% Remarkably,  Cooper-pair-like objects emerge at $\nu=2/5$, where holes attempt to attract each other despite Coulomb repulsion. This leads to possible superconductivity in Mott insulators \cite{andrei_rev}.
\begin{figure*}[t!]
\centering
   \includegraphics[width=1.9\columnwidth]{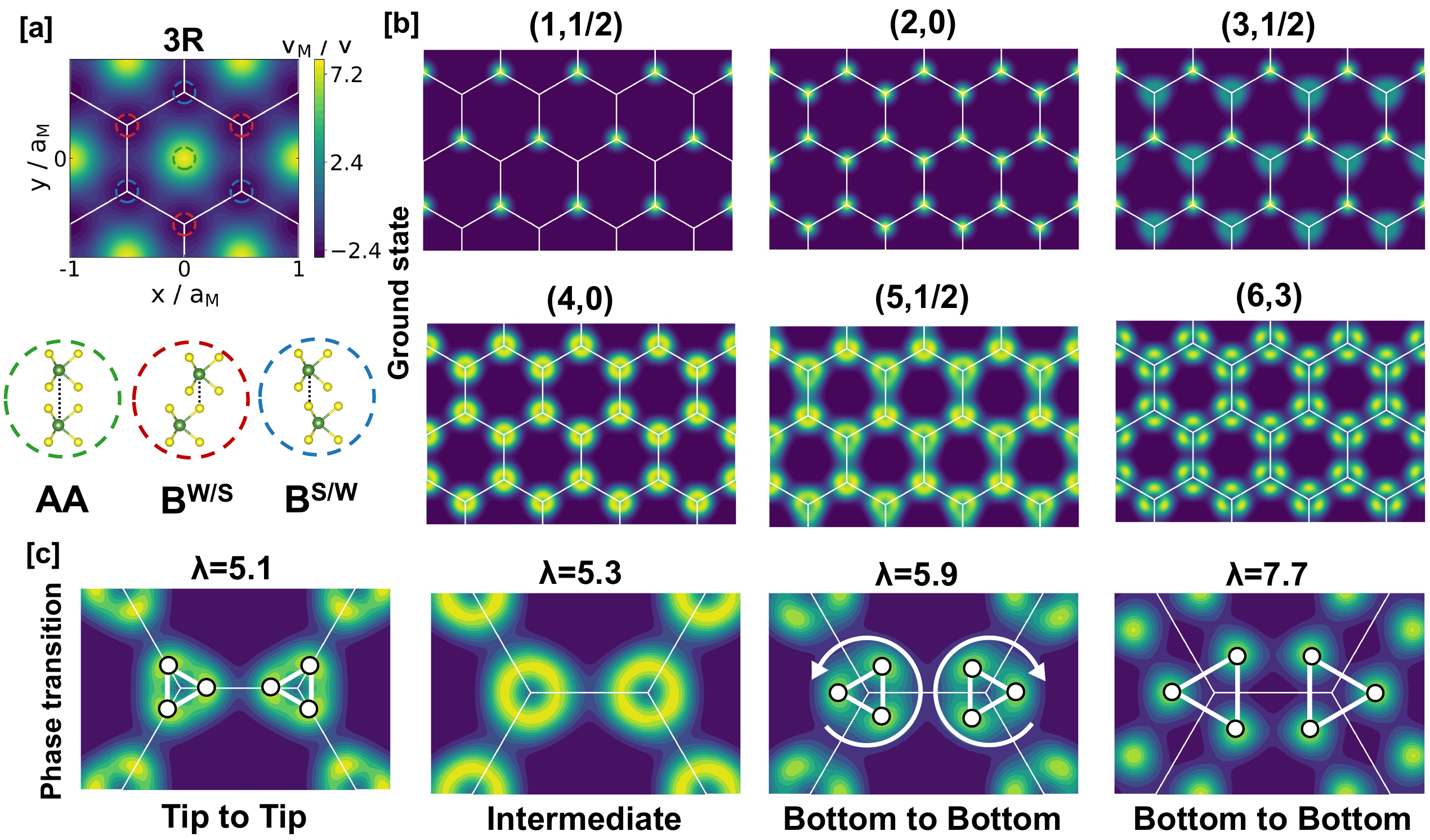}
    \caption{\textbf{Calculated density pattern of ${\rm 3R\ WS_2}$ homobilayer}. Tuples above the figure denote the particle fillings and the magnetic number of spin ($\nu$, $S_z$). \textbf{a}, moir\'e potential of ${\rm 3R\ WS_2}$ homobilayer. $a_M$ denotes length of moir\'e cell and is set to $16.51\ {\rm nm}$. \textbf{b}, calculated ground states at different particle fillings $\nu$. \textbf{c}, phase transition of $(6,3)$ state at different interaction ratio $\lambda$. $\lambda$ measures the strength ratio between Coulomb repulsion and moir\'e potential attraction. Moir\'e potential amplitude $V$ is tuned to control $\lambda$. The orientation of trimers get flipped with increasing $\lambda$ to avoid strong Coulomb repulsion.}
    \label{fig:3r}
\end{figure*}
\begin{figure*}[t]
\centering
   \includegraphics[width=2\columnwidth]{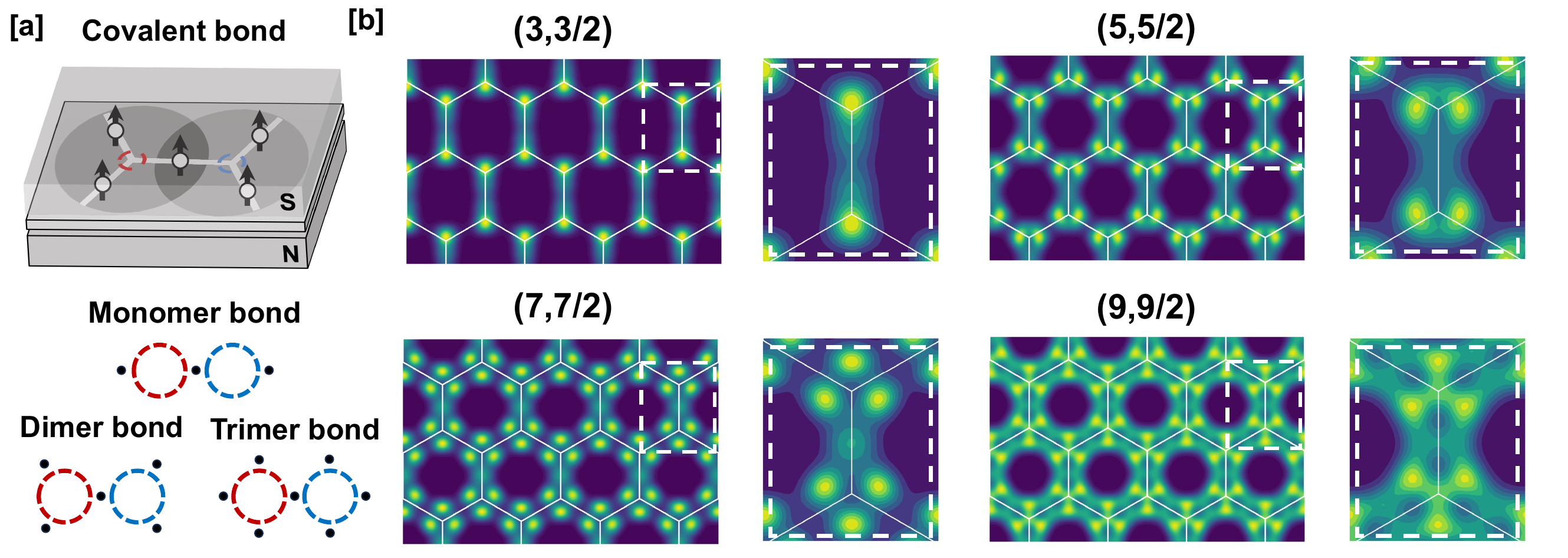}
    \caption{\textbf{Calculated covalent states of ${\rm 3R\ WS_2}$ homobilayer}. \textbf{a}, schematic plot of covalent bond formation in Wigner phases. Upper panel: formation of covalent bond in a spin polarized system. Lower panel: particle patterns of the monomer, dimer and trimer bond. Colored circles denote $\rm B^{W/S}$ and $\rm B^{S/W}$ regions of 3R $\rm WS_2$ respectively.
    \textbf{b}, particle density pattern of several calculated covalent states. $a_M$ is set to $16.51\ {\rm nm}$. Tuples above the figure denote the particle fillings and the magnetic number of spin ($\nu$, $S_z$). In each sub-panel, the right figure shows a zoom-in of the covalent bond.
    %$a_M$ is set to 16.51 nm.
    }
    \label{fig:covalent}
\end{figure*}

% \begin{figure*}[t]
% \centering
%    \includegraphics[width=2\columnwidth]{various_phase.pdf}
%     \caption{\textbf{Calculated patterns of ${\rm WS_2\ (3R)}$ of different quantum numbers}. Tuples above the figure denote the particle fillings and magnetic number of spin ($\nu$, $S_z$). Systems are set to $\theta=1.1^\circ$. }
%     \label{fig:various_phase}
% \end{figure*}

\paragraph{Wigner molecular crystal}
Beyond the aforementioned fractional filling, recent research has been extended to multiple integer fillings of moir\'e superlattice, wherein multiple particles are assigned to a single moir\'e cell \cite{wigner_molcule_theory, wigner_molecule_exp}. These particles aggregate and form a ``molecule", naturally residing at the minimum of the moir\'e potential. Upon the activation of Coulomb repulsion, these molecules slightly disperse and form the so-called Wigner molecular crystal \cite{wigner_molcule_theory}.
These molecular crystals are recently observed in ${\rm  2H\ WS_2}$ homobilayer via scanning tunneling microscopy (STM) \cite{wigner_molecule_exp}, prompting us to employ our neural network to explain this observation. 

Quantum states of ${\rm 2H\ WS_2}$ with different fillings $\nu$ and spin quantum numbers $S_z$ are investigated and the results are plotted in Fig.~\ref{fig:2h}. At $\nu=2$ fillings, we find $S_z=0$ state has the lowest energy in which two spin-opposite holes are all placed in AB region. 
If we align their spins parallel, one hole will depart due to Pauli exclusion and transfer to the ${\rm B^{S/S}}$ region, forming a charge-transfer insulator as an excited state \cite{charge_transfer}.
A more intricate situation happens at $\nu=3$.
The previous understanding of the experimental pattern at $\nu=3$ presumes the ground state to be fully spin-polarized \cite{wigner_molecule_exp}, while we find that the $S_z=1/2$ state presents a rather similar trimer pattern with $S_z=3/2$ and a slightly lower energy 0.4 meV / hole (Supplementary Fig.~2). 
In either trimer pattern, particles mainly accumulate at the vertexes of the triangle, while the center remains empty to minimize Coulomb interactions.  
More detailed experimental characterization is desired to resolve such a delicate competition and confirm the true ground state.
Concerning $\nu=4$, $S_z=1$ state is identified as the ground state, exhibiting a more hollow triangle pattern. It is also intriguing to see a flipping of the triangle when the system enters a fully polarized $S_z=2$ state, which can be further verified in experiment by applying a magnetic field. 
Overall, the predicted ground-state patterns from deep learning are perfectly consistent with recent STM observations at different fillings \cite{wigner_molecule_exp}, showing the reliability of our deep learning approach.
%It is also worth noting that the theoretical explanation of experimental pattern at $\nu=3$ presumes the ground state to be a fully spin-polarized state, but our simulation suggests that $S_z=1/2$ state is more likely to be the ground state observed in experiments (Extended Data Fig.XXX).

\paragraph{Predicting new phases}
%To date, only the Wigner molecular crystals of $C_3$ symmetry have been observed, leaving additional symmetry classes yet to be discovered. In this work, we employ our network to explore  $C_6$ symmetric Wigner molecular crystal in 3R ${\rm WS_2}$ homobilayer \cite{WS2_3R}.
Having demonstrated the capability of our deep learning approach to reproduce the experimentally observed phases, we now proceed to further explore new phases.
As an example, we examine the ${\rm WS_2}$ homobilayer with 3R configuration, which differs from 2H configuration due to the opposite orientations of layers \cite{WS2_3R}. Notably, the energies of different Bernal regions ($\rm B^{W/S}$ and $\rm B^{S/W}$) coincide with each other in 3R ${\rm WS_2}$, resulting in an underlying $C_6$ symmetry (see Fig.~\ref{fig:3r}a). 

The calculated ground states at different particle fillings are plotted in Fig.~\ref{fig:3r}b. The first interesting phenomenon we observe is the spontaneous symmetry breaking in the moir\'e pattern. At $\nu=1$, holes are located at corners of the honeycomb and leave the adjacent sites empty, which breaks $C_6$ symmetry to $C_3$ and minimizes Coulomb interactions. This broken $C_6$ symmetry will then get restored if doping one more hole. All moir\'e potential minima will be occupied by holes and make a perfect honeycomb lattice. As the doping increases to $\nu=3$, two spin-opposite holes are forced to accompany each other in one corner of the honeycomb despite the Coulomb repulsion, forming a more diffuse pattern than the adjacent singly occupied site. This leads to two distinct density patterns appearing in the sites of the honeycomb, breaking the $C_6$ symmetry to $C_3$, and the symmetry will again get restored at $\nu=4$ since all the sites are doubly occupied then. At $\nu=5$, a special density pattern appears, which exhibits a mirror symmetry. It is also worth noting that $(5,3/2)$ state are nearly degenerate with the calculated ground state $(5,1/2)$ while showing a distinct $C_3$ symmetry in Supplementary Figs.~3 and 4.

%As we push fillings $\nu$ to 6, an intriguing phase competition begins when two trimers formed in the sites of the honeycomb with two possible orientations (see Fig.~\ref{fig:3r}c), determined by the interaction ratio $\lambda\sim \epsilon^{-1} V^{-1/4}$ between Coulomb repulsion and moir\'e potential attractions \cite{wigner_molcule_theory}. 
When the filling $\nu$ reaches 6, trimers form in the sites of the honeycomb lattice. 
The trimers have different orientations and an intriguing phase transition can be induced by tuning either the Coulomb repulsion or the moir\'e potential attraction.
In Fig.~\ref{fig:3r}c, we plot the zoom-in pattern of the $(6,3)$ state at varying interactions, defined by the interaction ratio $\lambda\sim \epsilon^{-1} V^{-1/4}$ \cite{wigner_molcule_theory}.
When the moir\'e potential attraction dominates over the Coulomb repulsion, these two triangles get placed in a tip-to-tip manner, which resembles the shape of moir\'e potential. As the Coulomb repulsion gets stronger, triangles become larger and gradually melt into circles, finally getting flipped and forming a bottom-to-bottom pattern. Holes are then placed in the saddle points of moir\'e potential, which avoids strong Coulomb interaction compared to the tip-to-tip pattern.

\paragraph{Wigner covalent crystal} 
Up to now, most discovered Wigner phases appear as atomic crystals and molecular crystals, where ``atoms'' or ``molecules'' are isolated units that resemble chemical atoms and molecules.
%An interesting observation is that the real atoms and molecules can further form bonding, whereas similar features have not been observed in experiments of Wigner phases.
%
Interestingly, the equivalent of bonding commonly seen in molecules is notably absent in studies of Wigner phases.
Here, by tuning the filling number and spin polarization, we show that it is possible to form covalent bonding in moir\'e systems.
Fig.~\ref{fig:covalent}a shows a schematic plot of the formation mechanism of a covalent bond between two Wigner molecules. 
Assuming each molecule already has several holes forming a relatively stable state, then excess holes would feel repulsive interactions from the existing holes. They will be pushed towards the middle regime of two molecules and reach equilibrium.
This mechanism is similar to the ``sharing'' of electrons between two covalent bonded molecules. 
Such a mechanism can be further facilitated by enhancing Pauli exclusion, which is achievable by applying a magnetic field to align the spin of all holes in the same orientation (Fig.~\ref{fig:covalent}a).

Fig.~\ref{fig:covalent}b demonstrates the emergence of covalent bonds in fully polarized Wigner phases.
%In our simulations, we have identified a novel covalent crystal that displays interesting symmetries and mechanisms. 
%
%Our findings, as illustrated in Fig.~\ref{fig:3r}d, suggest that when one more hole is doped to $\nu=2$ states and gets fully polarized, the additional hole becomes diffused between the honeycomb sites, thus creating a "covalent bond" in the honeycomb edge.
%
%Consequently, the previously perfect honeycomb lattice of $\nu=2$ state deforms due to the Coulomb interaction and breaks $C_6$ symmetry to $C_2$.
%
%Our findings, as illustrated in Fig.~\ref{fig:covalent}, indicate that "covalent bonds" emerge at the honeycomb edge as holes diffuse between the honeycomb sites. 
%
Specifically, the first panel shows the formation of covalent bonds in a fully polarized $\nu=3$ state when an additional hole is doped into the $C_6$ symmetric $\nu=2$ state. 
Consequently, the honeycomb lattice of the $\nu=2$ state deforms due to Coulomb interactions, breaking its $C_6$ symmetry down to $C_2$.
Similar situations occur in polarized $\nu$ = 5 and 7 states, except that the honeycomb sites are now occupied by dimers and trimers respectively.
As we further increase the filling to $\nu=9$, honeycomb sites are connected in a network of covalent bonds, forming a novel crystal, which we dub ``Wigner covalent crystal'' following the existing names of Wigner crystal and Wigner molecular crystal.
The Wigner covalent phase adds another key member to the family of Wigner phases.   

\paragraph{Discussion}
Moir\'e materials host various novel phenomena, tightly related to strong correlations. Despite the rapid development of experimental observation, the simulation approach of these materials remains at an early stage, which hinders a thorough understanding of the strongly correlated phenomena.   
%
% Despite the fruitful phase has been discovered, more phases remain untouched due to computational and experimental limitations.
%In this work, we described a deep learning approach to simulate moir\'e materials. Numerous phase with different symmetry appears in our simulation, proving it to be a general, accurate and efficient framework to study moir\'e materials. 
%In the future, we believe our work can provide more moir\'e patterns for discover. Moreover, our method can be extended to other important problems, such as multi-layer materials, quantum transparant and anomalous Hall effect.  
%
In this work, we described a deep learning approach to simulate moir\'e materials with a rigorous treatment of electron correlations.
Our simulations reveal numerous exotic phases with varying symmetries, demonstrating the generality, accuracy, and efficiency of this approach for exploring moir\'e materials.
Besides providing novel moir\'e patterns, our framework can also be extended to predict more material behaviors and properties, bypassing certain experimental constraints. 
This work will help accelerate the discovery of new moir\'e materials, including multi-layer systems, and their applications in the future. 
Furthermore, our approach shows promise for broader research areas, such as the quantum transparency phenomenon and the anomalous Hall effect.

\section*{Methods}

\paragraph{Effective model of TMD}
Moir\'e materials are known to contain enormous atom numbers in the unit cell, which pose a great challenge for direct \textit{ab initio} simulations. Nonetheless, an effective model can be derived for doping holes in TMD materials to simplify the problem \cite{TMD_model,WS2_3R}. 

According to the density functional theory (DFT) result \cite{TMD_model,WS2_3R},  heterobilayers such as ${\rm WSe_2/WS_2}$ exhibit valence band maximums located at $K$ and $K'$ points of the Brillouin zone, which have the same energy but opposite spin due to spin-orbit coupling. On the other hand, homobilayers such as ${\rm WS_2}$ show valence band maximums located at $\Gamma$ point with spin degeneracy, as a result of Kramers' theorem. In either case, the most active valence states show a spin degeneracy, which can be treated as a quasi-particle with spin degrees as shown in Eq.~\eqref{eq:model}. Furthermore, the effective mass $m^*$ is fitted to describe the dispersion of DFT valence band edge and moir\'e potential is derived to mimic the interaction between doping holes and inertia electrons far below Fermi surface, see Refs.~\cite{2d_model, TMD_model, WS2_3R} for more detailed derivations. When the moir\'e system gets twisted to critical angles, flat bands appear indicating that the effective mass of the particle approaches infinity. Kinetic energy vanishes and Coulomb interaction dominates in the Hamiltonian, which requires high-accuracy wavefunction methods to incorporate correlations among particles. 

It's also worth noting that the moir\'e potential of 3R materials shows a slightly different form from Eq.~\eqref{eq:model} \cite{WS2_3R}, which reads
\begin{equation}
\begin{gathered}
    V^{\rm 3R}_M(\mathbf{r})=-2\sum_{i=1}^3\sum_{s=1}^3 V_s \cos(\mathbf{b_i^s}\cdot\mathbf{r}+\phi)\ , \\
    \mathbf{b}_i^1 = \frac{4\pi}{\sqrt{3}a_M}(\sin\frac{2\pi i}{3}, \cos\frac{2\pi i }{3})\ , \\
    \mathbf{b}_i^2 = \frac{4\pi}{a_M}\left[\sin(\frac{2\pi i}{3}+\frac{\pi}{6}), \cos(\frac{2\pi i }{3}+\frac{\pi}{6})\right]\ , \\
    \mathbf{b}_i^3 = \frac{8\pi}{\sqrt{3}a_M}(\sin\frac{2\pi i}{3}, \cos\frac{2\pi i }{3})\ , \\
\end{gathered}
\end{equation}
where $\mathbf{b}_i^s$ denote reciprocal vectors of different shells. Specific parameters of various materials are summarized in Supplementary Note 2. 

\paragraph{Neural network architecture}
Our neural network resembles similar architecture to previously proposed networks for solid materials and uniform electron gas \cite{deepsolid,ferminet_heg}.
%which will be reviewed briefly for self-consistency. 

Particle features $\mathbf{r}_i$ are combined with each other and form permutation equivalent features $\mathbf{f}_i$ \cite{ferminet}

\begin{gather*}
\mathbf{f}_i={\rm concat}\left[\mathbf{g}_i^S, \mathbf{g}_i^D\right],
 \\
 \mathbf{g}_i^S={\rm concat}\left[\sin(\mathbf{r}_i\cdot\mathbf{b})\ \mathbf{a}, \sum_j\sin(\mathbf{r}_j\cdot\mathbf{b})\ \mathbf{a}\right],\\
  \mathbf{g}_i^D={\rm concat}\left[\sum_j\sin[(\mathbf{r}_i-\mathbf{r}_j)\cdot\mathbf{b}]\ \mathbf{a}, 
  \sum_j d(\mathbf{r}_i-\mathbf{r}_j)\right].
\label{eq:feature}
\end{gather*}
These features are subsequently fed into a neural network to construct effective orbital functions $u_{\mathbf{k}_i}$ which capture correlations between particles
\begin{gather*}
     u_{\mathbf{k}_i}(\mathbf{r}_1,...,\mathbf{r}_N)={\rm Network}(\mathbf{f}_j)\ .
\label{eq:bloch}
\end{gather*}
%\begin{gather*}
%     e^{i\mathbf{k}_i\cdot\mathbf{r}_j}u_{\mathbf{k}_i}(\mathbf{r}_1,...,\mathbf{r}_N)={\rm Network}(\mathbf{f}_j).
%\label{eq:bloch}
%\end{gather*}
These orbitals are then combined with a momentum dependent phase factor $e^{i\mathbf{k}_i\cdot\mathbf{r}_j}$ to become elements of Slater determinants, which forms a legal antisymmetric many-body wavefunction for particles in moir\'e superlattice.
To ensure the periodic boundary condition, we employ the triangle distance input feature $d(\mathbf{r})$, which reads \cite{ferminet_heg}
 \begin{equation}
\begin{split}
4\pi^2 d^2(\mathbf{r})=&\sum_{ij}\sin(\mathbf{r}\cdot \mathbf{b}_i)\sin(\mathbf{r}\cdot \mathbf{b}_j)\ \mathbf{a}_i\cdot\mathbf{a}_j\\
&+[1-\cos(\mathbf{r}\cdot \mathbf{b}_i)][1-\cos(\mathbf{r}\cdot \mathbf{b}_j)]\ \mathbf{a}_i\cdot\mathbf{a}_j\ .\\
\end{split}
\label{eq:distance}
\end{equation}
%Other modifications include the changing from three dimension to two dimension and discarding the exponential decay term in neural networks. 
Specific hyperparameters of network are given in Supplementary Note 1.

\paragraph{Neural network optimization}
Considering the wide range of length scale in moir\'e superlattice, we rescale the effective model $\mathbf{r}\rightarrow a_M\mathbf{r}'$ for computational convenience, which reads
\begin{equation}
\begin{gathered}
\hat{H}'\equiv\sum_i\left[-\frac{\Delta_i'}{2}+m^*a_M^2V_M(\mathbf{r}_i')\right]+\frac{1}{2}\sum_{i\neq j}\frac{m^*a_M}{\epsilon|\mathbf{r}_i'-\mathbf{r}_j'|}.
\end{gathered}
\label{eq:scaled_model}
\end{equation}
Our networks is optimized to minimize the scaled Hamiltonian $\hat{H}'$, whose gradient reads
\begin{equation}
\begin{gathered}
\nabla\langle E_l\rangle=2\ {\rm Re}\left[\langle E_l\nabla\ln\Psi^*\rangle-\langle E_l\rangle\langle\nabla\ln\Psi^*\rangle\right], \\
E_l=\Psi^{-1}\hat{H}'\Psi,
\end{gathered}
\label{eq:gradient}
\end{equation}
and $\langle...\rangle$ denotes the expectations of operators with $|\Psi|^2$ distribution. Moreover, Kronecker factored curvature estimator (KFAC) optimizer \cite{kfac} is employed to train the neural network towards the ground states at different fillings and spin polarization.
%which significantly outperforms traditional optimizer in energy minimization. 

\paragraph{Moir\'e pattern analysis}
To visualize the Wigner phases, we plot the particle density $\rho(\mathbf{r})$ derived from many-body wavefunction $\Psi$ 
\begin{equation}
    \rho(\mathbf{r})=N\int d^3\mathbf{r}_2\cdots d^3\mathbf{r}_N \ |\Psi(\mathbf{r},\mathbf{r}_2,\cdots,\mathbf{r}_N)|^2\ .
    \label{eq:density}
\end{equation}
In practice, $\rho$ is evaluated by accumulating Monte Carlo samples of particles on a $100\times100$ uniform grid over the moir\'e cell. 

\paragraph{Workflow and computational details} Supercell approximation is employed in our simulations. Particles are initialized uniformly in the moir\'e supercell, and gradually form moir\'e patterns during energy minimization. The expectations of operators are evaluated via the Monte Carlo approach. Forward Laplacian technique is employed to speed up the simulation \cite{lapnet}. Most simulations in this work are performed on eight A800 graphics processing units within several hours. Training curves of each system are plotted in Supplementary Figs.~1-3. Other excited states are plotted in Supplementary Fig.~5. Calculated energy and geometry are listed in Supplementary Note 3.

\subsection*{Data availability}
The data generated in this study are provided in the Supplementary Information.

\subsection*{Code availability}
This work is developed upon open-source \href{https://github.com/bytedance/DeepSolid}{DeepSolid} \cite{deepsolid_code} on GitHub.

\begingroup
\setlength\bibsep{0pt}
\newcommand{\mathsl}{\mathit}
\footnotesize
\bibliography{reference}
\endgroup

% \bibliography{reference}% Produces the bibliography via BibTeX.

\subsection*{Acknowledgements}
\begingroup
\footnotesize
We want to thank Hongyuan Li for discussion. We want to thank ByteDance Research Group for inspiration and encouragement. This work is directed and supported by Hang Li and ByteDance Research. This work is supported by the National Key R\&D Program of China (2021YFA1400500 to J.C., 2021YFA1401300 to Y.X.) and the Strategic Priority Research Program of Chinese Academy of Sciences (XDB33000000).
\endgroup

\subsection*{Author contributions}

\begingroup
\footnotesize
X.L., Y.X. and J.C. conceived the study; X.L. developed the method, performed implementations, simulations and data analysis; Y.X. and J.C. supervised the project. X.L., Y.Q., W.R., Y.X., and J.C. wrote the paper.
\endgroup

\end{document}

\title{Supplementary information: Emergent Wigner phases in moir\'e superlattice from deep learning}

\maketitle
\section{Hyperparameters for simulations}

The recommended hyperparameters are listed in Supplementary Table~\ref{tab:hyper}.
\begin{table}[htb]
\centering
\caption{\textbf{Recommended hyperparameters}
}
\begin{tabular}{lclc}
\toprule
Hyperparameter & Value & Hyperparameter & Value  \\
\midrule
Dimension of one electron layer $\mathbf{V}$ & 256 & Dimension of two electron layer $\mathbf{W}$ & 32 \\
Number of layers  & 4 & Number of determinants & 8\\
Optimizer & KFAC & Learning rate & 1e-2\\
Damping & 1e-3 & Constrained norm of gradient & 1e-3 \\
Momentum of optimizer & 0.0 & Batch size & \num{4096} \\
Number of training steps & 5e4 & Clipping window of gradient & 5 \\
MCMC burn in & 1e3 & MCMC steps between each iterations & 30 \\
MCMC move width & 2e-2 & Target MCMC acceptance & 55\% \\
Precision & Float32 &  Number of inference steps & 1e4 \\  
\bottomrule
\end{tabular}
\label{tab:hyper}
\end{table}

\section{Parameter of moir\'e potential}
% %
% Moire potential of ${\rm WSe_2/WS2}$ and ${\rm WS_2\ (2H)}$ reads
% \begin{equation}
% \begin{gathered}
%     V_M(\mathbf{r})=-2V\sum_{i=1}^3 \cos(\mathbf{b_i}\cdot\mathbf{r}+\phi), \\
%     \mathbf{b}_i = \frac{4\pi}{\sqrt{3}a_M}(\sin\frac{2\pi i}{3}, \cos\frac{2\pi i }{3}).
% \end{gathered}
% \end{equation}
Employed model parameters of ${\rm WSe_2/WS_2}$ and ${\rm 2H\ WS_2}$ are taken from Refs.~\cite{wigner_molecule_exp, charge_transfer} and listed in Supplementary Table~\ref{tab:moire_1}.
\begin{table}[htb]
\centering
\caption{\textbf{Model parameters of ${\rm WSe_2/WS_2}$ and ${\rm 2H\ WS_2}$}
.}
\begin{tabular}{lcccc}
\toprule
Systems & V / meV & $m^*/m_e$ & $\phi$ & $\epsilon$   \\
\midrule
${\rm WSe_2/WS_2}$ & 15 & 0.5 & 45$^\circ$ & 5 \\
${\rm 2H\ WS_2}$ & 10.3 & 0.9 & 20$^\circ$ & 5 \\
\bottomrule
\end{tabular}
\label{tab:moire_1}
\end{table}

Model of 3R ${\rm WS_2}$ has a slightly different form and its parameters are listed in Supplementary Table~\ref{tab:moire_2} \cite{WS2_3R}.
\begin{table}[htb]
\centering
\caption{\textbf{Model parameters of ${\rm 3R\ WS_2}$}
}
\begin{tabular}{lcccccc}
\toprule
Systems & $V_1$ / meV & $V_2$ / meV & $V_3$ / meV & $m^*/m_e$ & $\phi$ & $\epsilon$   \\
\midrule
${\rm 3R\ WS_2}$ & 33.5 & 4.0 & 5.5& 0.87& 180$^\circ$ & 5 \\
\bottomrule
\end{tabular}
\label{tab:moire_2}
\end{table}

\section{Calculated energy and geometry}
Calculated energies and geometry of each system are listed in Supplementary Table~\ref{tab:energy_geo}.
\begin{table}[htb]
\centering
\caption{\textbf{Calculated energies and geometry of each system}} 
\begin{tabular}{lcccccc}
\toprule
Systems & $a_M/\ {\rm nm}$ & fillings $\nu$& $S_z$&energy per hole / meV  & lattice vectors / $a_M$ & supercell size  \\
\midrule
${\rm WSe_2/WS_2}$ & 8.3 & 1 & 1 &  -80.2965(2) & $a_1=(1,0),\ a_2=(1/2, \sqrt{3}/2)$ & $5\times5$ \\
 &  & 2/3 & 1 &  -67.0221(1) & $a_1=(3/2,\sqrt{3}/2),\ a_2=(-3/2, \sqrt{3}/2)$ & $4\times4$ \\
 &  & 1/3 & 1/2 &  -53.97697(9) & $a_1=(1,0),\ a_2=(1/2, \sqrt{3}/2)$ & $5\times5$\\
 &  & 1/2 & 1 &  -59.39885(9) & $a_1=(2,0),\ a_2=(1, \sqrt{3})$ & $4\times4$ \\
 &  & 2/5 & 5 &  -55.7933(2) & $a_1=(5,0),\ a_2=(5/2, 5\sqrt{3}/2)$ & $2\times2$\\
 &  & 1/6 & 3 &  -42.41933(9) & $a_1=(6,0),\ a_2=(3, 3\sqrt{3})$ & $2\times2$\\
\midrule
${\rm 2H\ WS_2}$ & 9.8 & 2 & 0 & -84.3421(2)  & $a_1=(1,0),\ a_2=(1/2, \sqrt{3}/2)$ & $3\times3$ \\
 &  & 2 & 1 & -81.64466(7) & & \\
\midrule
${\rm 2H\ WS_2}$ & 9.8 & 3 & 1/2 & -90.6866(2) & $a_1=(1,0),\ a_2=(1/2, \sqrt{3}/2)$ & $4\times4$\\
 &  & 3 & 3/2 & -90.2980(1) \\
\midrule
${\rm 2H\ WS_2}$ & 9.8 & 4 & 0 & -98.2843(3) & $a_1=(1,0),\ a_2=(1/2, \sqrt{3}/2)$ & $4\times4$ \\
 &  & 4 & 1 & -98.4657(2) \\
 &  & 4 & 2 & -97.2430(1) \\
\midrule
${\rm 3R\ WS_2}$ & 16.51 & 1 & 1/2 & -102.51681(7) & $a_1=(1,0),\ a_2=(1/2, \sqrt{3}/2)$ & $3\times3$ \\
 &  & 2 & 1 & -114.6824(2) \\
 &  & 6 & 3 & -110.5943(2) \\
 &  & 7 & 7/2 & -110.9851(1) \\
 &  & 9 & 9/2 & -113.4288(2) \\
 \midrule
${\rm 3R\ WS_2}$ & 16.51 & 3 & 1/2 & -109.0860(1) & $a_1=(1,0),\ a_2=(1/2, \sqrt{3}/2)$ & $3\times3$ \\
 &  & 3 & 3/2 & -107.58514(8) \\
 \midrule
 &  & 4 & 0 & -110.2786(2) \\
 &  & 4 & 1 & -108.8960(1) \\
 &  & 4 & 2 & -107.8524(1) \\
 \midrule
 &  & 5 & 1/2 & -109.5465(1) \\
 &  & 5 & 3/2 & -109.54138(9) \\
 &  & 5 & 5/2 & -108.56062(6) \\
% \midrule
% ${\rm WS_2\ (3R)}$ & 2.5 & 3 & 1/2 & -227.11(4) & $a_1=(1,0),\ a_2=(1/2, \sqrt{3}/2)$ & $3\times3$\\
%  &  & 3 & 3/2 & -190.7439(5) \\
%
\bottomrule
\end{tabular}
\label{tab:energy_geo}
\end{table}

% Primitive cell lattice vectors and supercell sizes are listed in Supplementary Table~\ref{tab:geo}.
% %
% \begin{table}[htb]
% \centering
% \caption{\textbf{Geometry of primitive cell and supercell of each systems}
% }
% \begin{tabular}{lcccc}
% \toprule
% Systems & fillings $\nu$ & lattice vectors / $a_M$ & supercell size  & number of holes\\
% \midrule
% ${\rm WSe_2/WS_2}$ &  1& $a_1=(1,0),\ a_2=(1/2, \sqrt{3}/2)$ & $5\times5$ & 25\\
% ${\rm WSe_2/WS_2}$ &  2/3& $a_1=(3/2,\sqrt{3}/2),\ a_2=(-3/2, \sqrt{3}/2)$ & $4\times4$ & 32 \\
% ${\rm WSe_2/WS_2}$ &  1/3& $a_1=(1,0),\ a_2=(1/2, \sqrt{3}/2)$ & $5\times5$ & 25\\
% ${\rm WSe_2/WS_2}$ &  1/2& $a_1=(2,0),\ a_2=(1, \sqrt{3})$ & $4\times4$ & 32\\
% ${\rm WSe_2/WS_2}$ &  2/5& $a_1=(5,0),\ a_2=(5/2, 5\sqrt{3}/2)$ & $2\times2$ & 40 \\
% ${\rm WSe_2/WS_2}$ &  1/6& $a_1=(6,0),\ a_2=(3, 3\sqrt{3})$ & $2\times2$ & 24\\
% \midrule
% ${\rm WS_2\ (2H)}$ &  2 & $a_1=(1,0),\ a_2=(1/2, \sqrt{3}/2)$ & $3\times3$ & 18\\
% ${\rm WS_2\ (2H)}$ &  3 & $a_1=(1,0),\ a_2=(1/2, \sqrt{3}/2)$ & $4\times4$ & 48\\
% ${\rm WS_2\ (2H)}$ &  4 & $a_1=(1,0),\ a_2=(1/2, \sqrt{3}/2)$ & $4\times4$ & 64\\
% \midrule
% ${\rm WS_2\ (3R)}$ &  2 & $a_1=(1,0),\ a_2=(1/2, \sqrt{3}/2)$ & $3\times3$ & 18\\
% ${\rm WS_2\ (3R)}$ &  3 & $a_1=(1,0),\ a_2=(1/2, \sqrt{3}/2)$ & $3\times3$ & 27\\
% ${\rm WS_2\ (3R)}$ &  6 & $a_1=(1,0),\ a_2=(1/2, \sqrt{3}/2)$ & $3\times3$ & 54\\
% ${\rm WS_2\ (3R)}$ &  9 & $a_1=(1,0),\ a_2=(1/2, \sqrt{3}/2)$ & $3\times3$ & 81\\
% %
% \bottomrule
% \end{tabular}
% \label{tab:geo}
% \end{table}

\section{Training curves}
Training curves of each system are plotted in Supplementary Figs.~\labelcref{fig:train1,fig:train2,fig:train3}.
\begin{figure*}[t!]
    \centering
    \includegraphics[width=0.95\columnwidth]{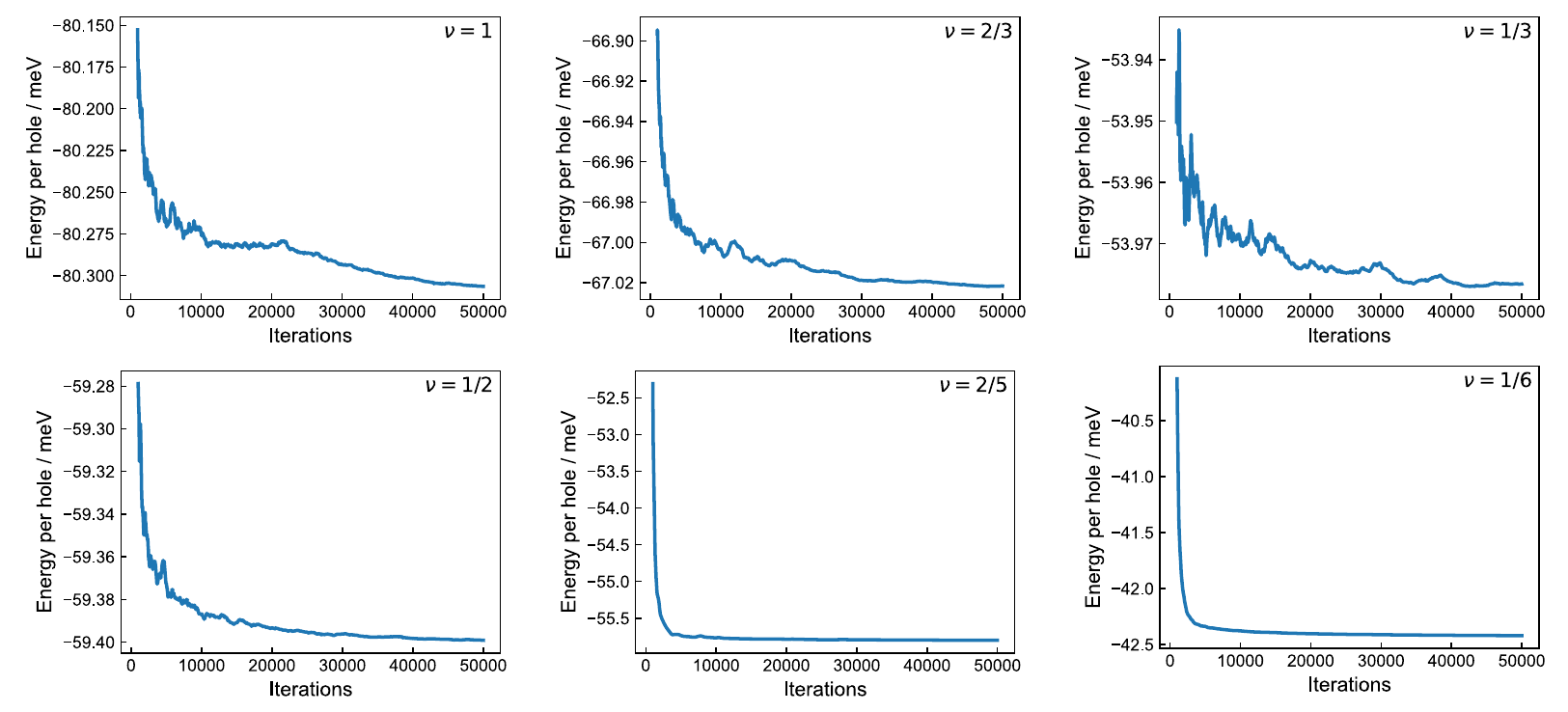}
    \caption{\textbf{Training curve of various states in ${\rm WSe_2/WS_2}$}. }
    \label{fig:train1}
\end{figure*}

\begin{figure*}[t!]
    \centering
    \includegraphics[width=0.95\columnwidth]{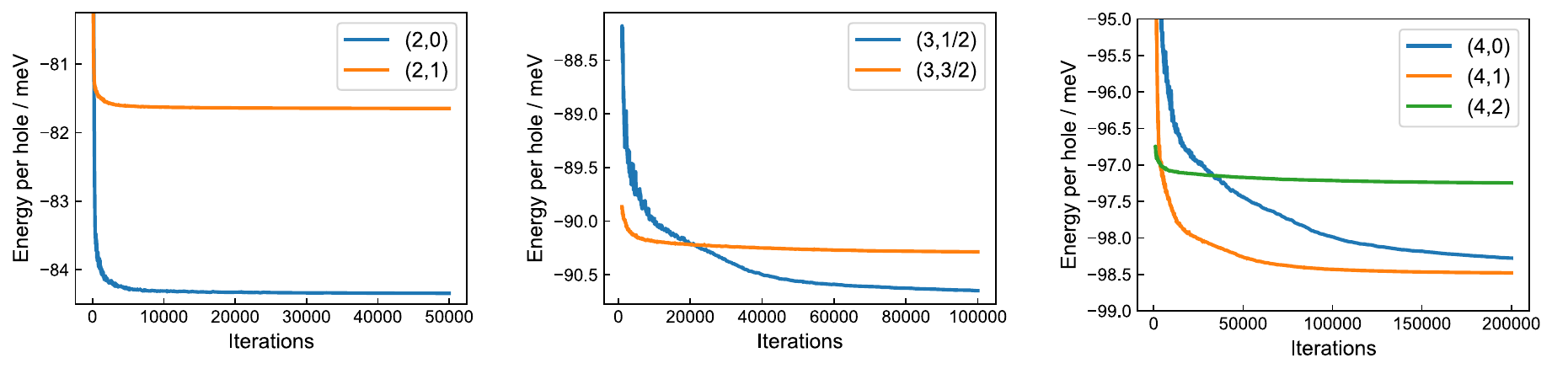}
    \caption{\textbf{Training curve of various states in ${\rm 2H\ WS_2}$}. Tuples in the legends denote the particle fillings and the magnetic number of spin ($\nu$, $S_z$).}
    \label{fig:train2}
\end{figure*}

\begin{figure*}[t!]
    \centering
    \includegraphics[width=0.95\columnwidth]{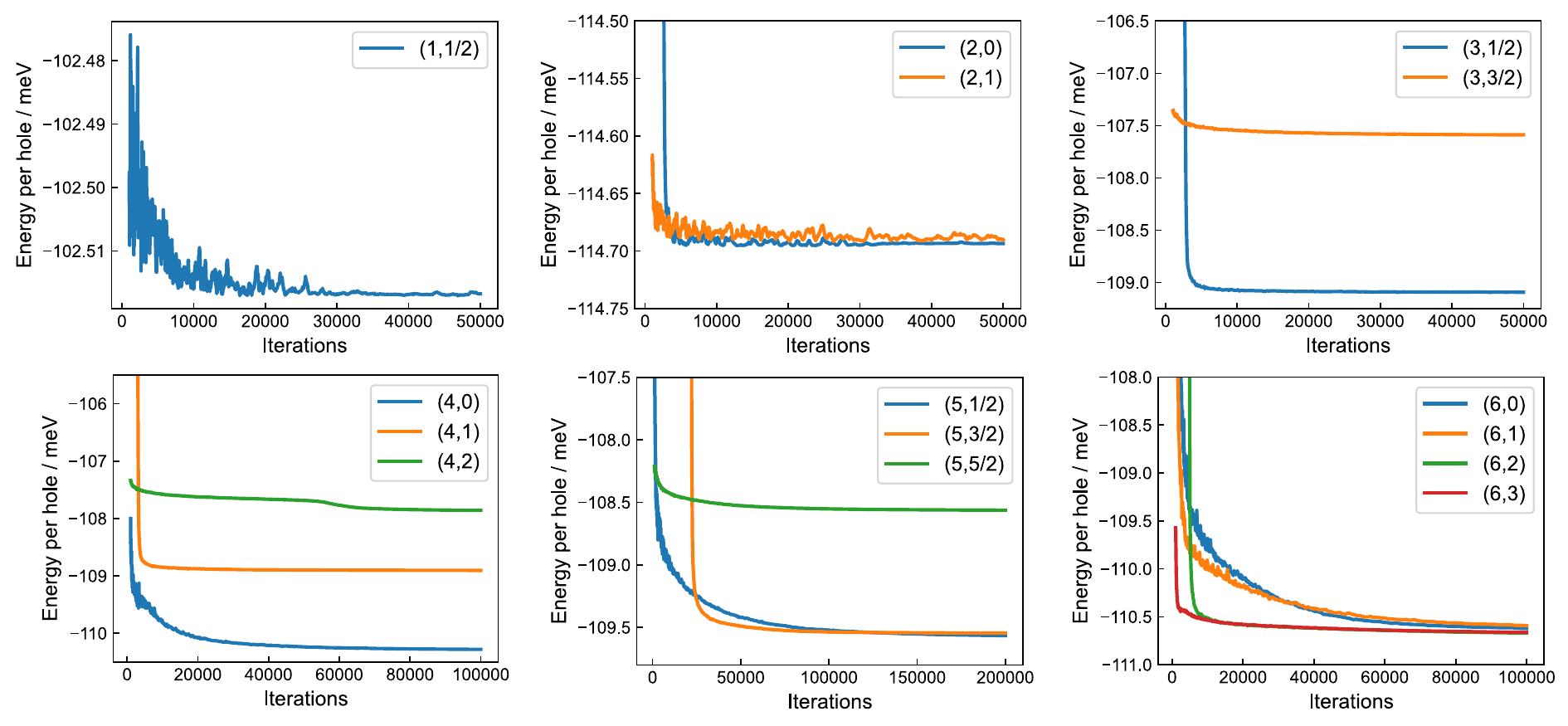}
    \caption{\textbf{Training curve of various states in ${\rm 3R\ WS_2}$}. Tuples in the legends denote the particle fillings and the magnetic number of spin ($\nu$, $S_z$).}
\label{fig:train3}
\end{figure*}

\section{Other states}
\begin{figure*}[t!]
    \centering
    \includegraphics[width=1\columnwidth]{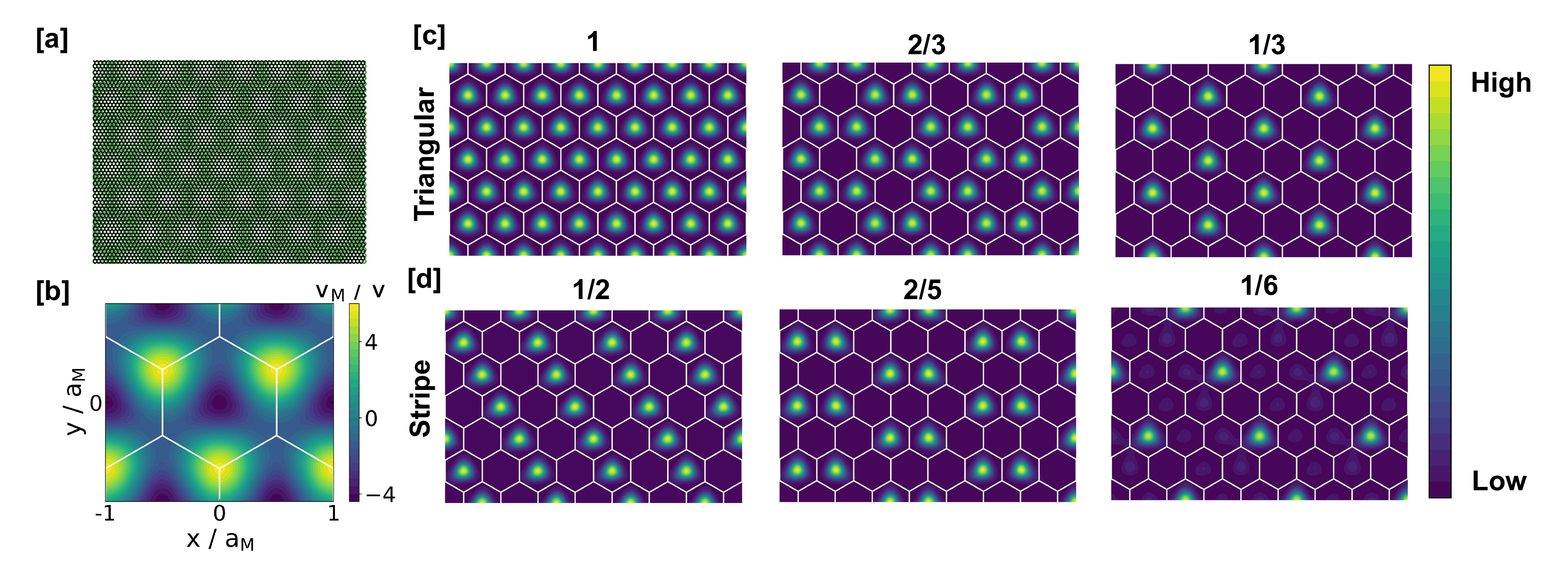}
    \caption{\textbf{Calculated hole density pattern of ${\rm WSe_2/WS_2}$ heterobilayer}. Numbers above the figures denote the particle fillings $\nu$. \textbf{a}, structure of the ${\rm WSe_2/WS_2}$ superlattice. \textbf{b}, moir\'e potential of ${\rm WSe_2/WS_2}$. $a_M$ denotes corresponding length of moir\'e cell and is set to ${\rm 8.3\ nm}$. \textbf{c}, calculated triangular phase at fillings $\nu=1,2/3,1/3$. \textbf{d}, calculated stripe phase at fillings $\nu=1/2,2/5,1/6$.}
    \label{fig:frac}
\end{figure*}

\begin{figure*}[t!]
    \centering
    \includegraphics[width=1\columnwidth]{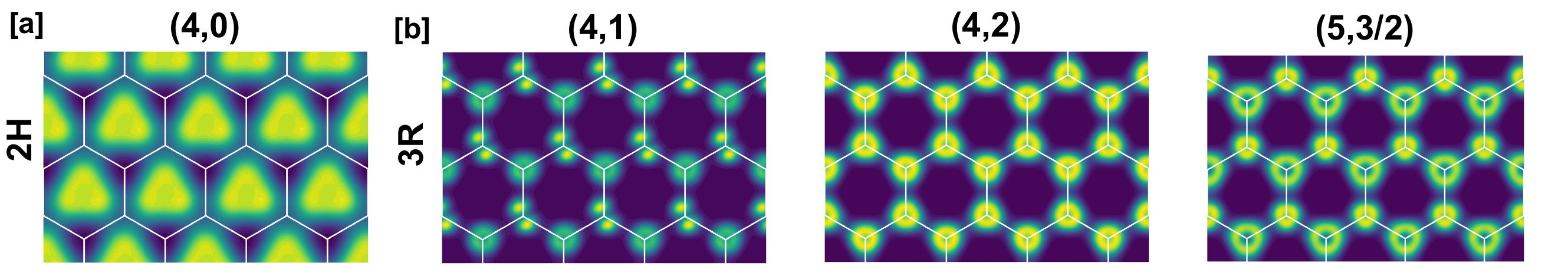}
    \caption{\textbf{Other density patterns of the ${\rm WS_2}$ system}. Tuples above the figure denote the particle fillings and the spin quantum number ($\nu$, $S_z$). \textbf{a-b}, moir\'e pattern of 2H and 3R ${\rm WS_2}$ respectively. All other states with $\nu=$ 2 and 6 of the 3R ${\rm WS_2}$ system show the same moir\'e pattern regardless of $S_z$.}
    \label{fig:other}
\end{figure*}

Generalized Wigner crystal states and other excited states are plotted in Figs.~\ref{fig:frac} and \ref{fig:other}.

\renewcommand{\refname}{Supplementary References}
\bibliography{supplement}
%\include{output.bbl}
%\bibliography{ref.bib}